\documentclass[twocolumn,prl]{revtex4}

\usepackage{graphicx}% Include figure files
\usepackage{dcolumn}% Align table columns on decimal point
\usepackage{bm}% bold math

\begin{document}

\title{Cosmological Magnetic Fields from Primordial Helical Seeds}

\author{G\"unter Sigl}
\affiliation{Institut d'Astrophysique de Paris, CNRS,
98bis boulevard Arago, 75014 Paris, France}

\begin{abstract}
Most early Universe scenarios predict negligible magnetic fields
on cosmological scales if they are unprocessed during
subsequent expansion of the Universe. We present a new numerical
treatment of the evolution of primordial fields and apply it to
weakly helical seeds as they occur in certain early Universe
scenarios. We find that initial helicities not much
larger than the baryon to photon number can lead to fields of
$\sim10^{-13}\,$G with coherence scales slightly below a
kiloparsec today.

\end{abstract}

%\pacs{PACS numbers: }

\maketitle

\section{Introduction}
The origin of galactic and large scale extragalactic magnetic fields
(for which there is no detection yet on scales larger than
mega-parsecs) is one of the main unresolved problems of astrophysics and
cosmology~\cite{review}.
In most scenarios where magnetic fields are produced
in the early Universe, these seed fields are concentrated
on scales below the horizon scale where they dissipate quickly, and
are too small on cosmological scales to have any observable
effects. However, if pseudoscalar interactions induce
a non-vanishing helicity of these seeds, such as in
string cosmology~\cite{string} or during the electroweak
phase transition by projection of non-abelian
Chern-Simons number onto the electromagnetic gauge
group~\cite{ew,cornwall,vachaspati}, then part of
the small scale power can cascade to large scales
and produce observable effects~\cite{cascade,cornwall,son,fc}.
In this paper we develop a new numerical approach to treat
such non-linear cascades up to zero redshift and apply it
to helical seed fields produced in the early Universe.

\section{MHD in the Early Universe}
The principal equations for magnetic field ${\bf B}$ and
velocity field ${\bf v}$ in the one-fluid approximation
of magnetohydrodynamics (MHD) are~\cite{choudhuri}
\begin{eqnarray}
  \partial_t{\bf B}&=&{\bf\nabla}\times({\bf v}\times{\bf B}-
  \eta\,{\bf\nabla}\times{\bf B})\,\nonumber\\
  \partial_t{\bf v}+({\bf v}\cdot{\bf\nabla}){\bf v}&=&
  \frac{({\bf\nabla}\times{\bf B})\times{\bf B}}{4\pi\rho}
  \,,\label{mhd}
\end{eqnarray}
where $\eta$ is the resistivity and $\rho$ is the fluid
density. The second equation describes
backreaction of the magnetic field on the flow. To
eliminate it, following Ref.~\cite{subra,cornwall}, we write
\begin{equation}
  {\bf v}\sim\frac{{\bf f}}{\beta\rho}=\frac{\tau}{4\pi\rho}
  ({\bf\nabla}\times{\bf B})\times{\bf B}\,,\label{vapprox}
\end{equation}
where $\beta$ is the drag coefficient and $\tau\equiv1/\beta$
is the fluid response time to the Lorentz force
${\bf f}=\left[({\bf\nabla}\times{\bf B})\times{\bf B}\right]/(4\pi)$.
The latter can be viewed as the time the charged fluid
can be accelerated until it interacts (scatters) with
other particles in the background and therefore describes
damping of the magnetic field modes.

Again following Ref.~\cite{subra}, we express the magnetic
field in terms of
correlation functions $M_{ij}(r,t)=\left\langle B_i({\bf x},t)
B_j({\bf y},t)\right\rangle$, where $r=|{\bf x}-{\bf y}|$,
assuming isotropy and homogeneity,
\begin{equation}
  M_{ij}=M_N\left(\delta_{ij}-\frac{r_i r_j}{r^2}\right)
  +M_L\frac{r_ir_j}{r^2}+H\epsilon_{ijk}r_k\,\label{mij}
\end{equation}
where $M_L$, $M_N$, and $H$ are longitudinal, transverse, and
helical magnetic correlation functions, respectively.
$M_N=\partial_r(r^2M_L)/(2r)$ is not independent because
of ${\bf\nabla}\cdot{\bf B}=0$. We define the magnetic
field and gauge invariant helicity power spectra per
logarithmic wavenumber interval $b^2(k)$ and
$h(k)$ by $E_M\equiv\left\langle{\bf B}^2({\bf r})\right\rangle
/(8\pi)=\int_{0}^{+\infty} dk b^2(k)/(8\pi k)$
and $H_M\equiv\left\langle{\bf B}({\bf r})\cdot{\bf A}({\bf r})\right\rangle
=\int_{0}^{+\infty} dk h(k)/k$, with ${\bf A}$ the vector
potential, ${\bf B}={\bf\nabla}\times{\bf A}$.
One can show that $M_L$ and $H$ are related to these
power spectra via
\begin{eqnarray}
  M_L(r)&=&
  \int_{0}^{+\infty} \frac{dk}{k}\frac{j_1(kr)}{kr}
  b^2(k)\nonumber\\
  H(r)&=&-\frac{1}{3r}\int_{0}^{+\infty}
  dk\,j_1(kr)h(k)\,,\label{kspace}
\end{eqnarray}
where $j_1(x)=\sin(x)/x^2-\cos(x)/x$ is the first order
spherical Bessel function. In terms of the usual
Fourier transforms ${\bf B}({\bf k})=\int d^3{\bf r}/(2\pi)^{3/2}
\exp(i{\bf k}\cdot{\bf r}){\bf B}({\bf r})$ etc.,
$b^2(k)=4\pi k^3{\bf B}^2(k)$ and $h(k)=k^3\int d\Omega_{\bf k}
{\bf B}({\bf k})\cdot{\bf A}({\bf k})$. Eq.~(\ref{kspace})
also shows that $M_L(0)=8\pi E_M/3$, and $H_M=-3\int_{0}^{+\infty}
drrH(r)$, and $|h(k)|\leq b^2(k)/k$ implies for all $r$
\begin{equation}
 |H(r)|\lesssim|M_L(r)|/r\equiv H_{\rm max}(r)\,.\label{hmax}
\end{equation}

Cosmological expansion can be taken into account by
redefining $M_L\to M_L/T^4$ and $H\to H/T^5$ from now on,
where $T$ is the cosmological temperature.
Assuming for now the absence of any external source
terms such as fluid motions 
except the one induced by the magnetic field, i.e.
using Eq.~(\ref{vapprox}), the MHD equations~(\ref{mhd})
reduce to
\begin{eqnarray}
\partial_t M_L&=&\frac{2}{r^4}\partial_r
\left(r^4\kappa\partial_r M_L\right) - 4\alpha TH\nonumber\\
\partial_t H&=&\frac{1}{r^4}\partial_r
\left[r^4\partial_r\left(2\kappa H+\alpha M_L/T\right)\right]
\label{mhd1}
\end{eqnarray}
where
\begin{eqnarray}
\kappa&=&\eta + \frac{\tau T^4}{2\pi\rho}M_L(0,t)\nonumber\\ 
\alpha&=&-\frac{\tau T^5}{\pi\rho}H(0,t)\,,\label{coeff}
\end{eqnarray}
and all quantities appearing here are in physical (not
co-moving) coordinates. The diffusion term $\kappa$ consists
of a microscopic ($\eta$) and a non-linear drift contribution,
whereas the $\alpha$ effect is only due to non-linear drift
here. The source terms will be discussed in the next section.
If the spatial derivatives
of $M_L$ and $H$ fall off faster than $1/r$ for $r\to\infty$,
Eq.~(\ref{mhd1}) implies $\partial_t H_M=9\left[2\kappa H(0)+
\alpha M_L(0)\right]$ which, together with Eq.~(\ref{coeff}),
shows that helicity is conserved in the absence of resistivity.

Eqs.~(\ref{mhd1}) describe small and large scale dynamos
of helical magnetic fields including damping by Ohmic
dissipation and "Silk" damping (which is expressed by
the redshift dependent relaxation time $\tau$)
on a unified basis. In the early Universe
the resistivity can be estimated
by $\eta\simeq1/(40\pi T)$ before photon decoupling,
$T\gtrsim0.25\,$eV~\cite{ae}, and by the Spitzer
resistivity $\eta\simeq\pi m_e^{1/2}e^2/T_e^{3/2}$ (where
$m_e$, $e$, and $T_e\sim10^6\,$K are electron mass,
charge, and temperature, assuming full ionization)
after recombination~\cite{choudhuri} (the results are
insensitive to the latter). Below $e^+e^-$ annihilation
at $T\simeq20\,$keV, within the MHD one fluid approximation
the fluid coupled to the magnetic field is well represented by the
tightly coupled remaining free electrons and protons and
$\tau$ is governed by Thomson scattering of photons off
electrons. In this regime we use~\cite{jko1}
\begin{eqnarray}
  \tau&=&\tau_\gamma\simeq4\times10^{21}
  \left(\frac{0.25{\rm eV}}{T}\right)^4\,X_e^{-1}\,{\rm cm}\nonumber\\
  \frac{\rho}{T^4}&\simeq&0.4\frac{4\pi^2}{45}
  \left(\frac{0.25{\rm eV}}{T}\right)
  \left(\frac{\Omega_bh^2}{0.0125}\right)\,,\label{taugamma}
\end{eqnarray}
where the number of free electrons per nucleon $X_e$
is $\simeq1$ for $T\gtrsim0.25\,$eV and $\simeq10^{-5}$
for $T\lesssim0.25\,$eV, and $\Omega_bh^2$ is the baryon
density in terms of the critical density times the
Hubble constant in units of $100\,{\rm km}/{\rm s}/{\rm Mpc}$
today. For $T\gtrsim20\,$keV we can approximate the
fluid to consist of the electromagnetically interacting
particles and $\tau$ is governed by neutrino scattering
with~\cite{jko1}
\begin{equation}
  \tau=\tau_\nu\simeq10^{11}
  \left(\frac{{\rm MeV}}{T}\right)^5
  \frac{8.75}{g_r-2}\frac{g_r}{g_\nu}\,{\rm cm}\,,\label{taunu}
\end{equation}
and $\rho/T^4\simeq g_f\pi^2/30$, 
where $g_r$, $g_f$, and $g_\nu=5.25$ are the statistical
weights of all relativistic particles, the particles
in the fluid and of the neutrinos, respectively.

\section{Helical Seeds}
Here we consider helical fluid motion, as it can arise during
cosmological phase transitions (see, for example, Ref.~\cite{ew,cornwall}
for the electroweak phase transition).
This case has already been treated in Ref.~\cite{subra}
which we adapt here to our situation. Since the backreaction
of the magnetic field onto the fluid motion has already
been taken into account by the approximation Eq.~(\ref{vapprox}),
the external fluid flow ${\bf v}_e$ is assumed to be
uncorrelated with the field. It is furthermore assumed that
the correlation time of the external velocity field is
much smaller than the time scale of change of the
magnetic correlation function, $\left\langle{\bf v}_{ei}({\bf x},t)
{\bf v}_{ej}({\bf y},s)\right\rangle=T_{ij}(r)
\delta(t-s)$, where, in analogy to Eq.~(\ref{mij}),
$r=|{\bf x}-{\bf y}|$, and
\begin{equation}
  T_{ij}=T_N\left(\delta_{ij}-\frac{r_i r_j}{r^2}\right)
  +T_L\frac{r_ir_j}{r^2}+C\epsilon_{ijk}r_k\,.\label{tij}
\end{equation}
Assuming for simplicity an incompressible fluid,
${\bf\nabla}\cdot{\bf v}_e=0$, the additional terms in
Eqs.~(\ref{mhd1}) and~(\ref{coeff}) are given by
\begin{eqnarray}
  \partial_t M_L&=&\cdots-\left(2\partial_r^2T_L+\frac{8}{r}
  \partial_r T_L\right)M_L\nonumber\\
  \kappa&=&\cdots+T_L(0)-T_L(r)\label{source}\\
  \alpha&=&\cdots+2C(0)-2C(r)\,,\nonumber
\end{eqnarray}
such that $\kappa$ and $\alpha$ obtain a scale dependent
turbulent diffusion and $\alpha$ effect contributions, respectively,
from the fluid. Here $T_L$ and $C$ are given by
$T_L(r)=\tau_{\rm corr}\left\langle{\bf r}\cdot{\bf v}_e(0)
{\bf r}\cdot{\bf v}_e({\bf r})\right\rangle/r^2$
and $C(r)=\tau_{\rm corr}\left\langle{\bf r}\cdot{\bf v}_e(0)
\times{\bf v}_e({\bf r})\right\rangle/(2r^2)$.
The correlation time $\tau_{\rm corr}$ is either
the damping time scale due to interactions with the background
or, if all components are tightly coupled and move as a whole,
the age of the Universe $t_u(T)$ at the relevant epochs.

The spatial velocity correlations $T_L$ and $C$ can be
expressed in terms of their power spectra $v^2(k)$
and $c(k)$, respectively, in complete analogy to Eq.~(\ref{kspace}).
In general they will have the form
\begin{eqnarray}
  T_L(r)&=&\frac{1}{3}\tau_{\rm corr}\left\langle v_e^2\right\rangle(T)
          f(r)\nonumber\\
  |C(r)|&\lesssim&|T_L(r)|/r\equiv C_{\rm max}(r)\,,\label{vparam}
\end{eqnarray}
where $f(r)$ is a dimensionless function with $f(r)\to1$ for $r\to0$
and, typically, a power law fall-off at large distances, and
the total power $\left\langle v_e^2\right\rangle(T)$
typically peaks at a certain temperature $T_{\rm ph}$,
for example, at a primordial phase
transition, and becomes negligible for $T\gg T_{\rm ph}$
and $T\ll T_{\rm ph}$.

\section{Numerical Simulations and Results}
For any early Universe scenario the initial conditions
for $M_L$ and $H$ at the temperature where the fields are
created should be calculated from the power spectra
$b^2(k)$ and $h(k)$, using Eq.~(\ref{kspace}). The
magnetic field evolution can then be obtained by numerically
integrating the non-linear partial differential Eqs.~(\ref{mhd1})
and their extensions with helical source terms in co-moving
coordinates from this initial time up to redshift zero. This is done by
employing an alternating implicit method~\cite{recipes}
to a one-dimensional grid of typically a hundred bins
roughly logarithmic in co-moving distance between the inverse
of today's cosmic microwave background (CMB)
temperature $T_0$ and $\sim10^4\,$Mpc, and using the logarithm of the
temperature $\ln T$ as independent variable, adopting the
standard relations between time and temperature, see
e.g. Ref.~\cite{kt}. In order to assure that induced velocities
Eq.~(\ref{vapprox}) remain non-relativistic, the induced contributions to the
coefficients $\kappa$ and $\alpha$, Eq.~(\ref{coeff}),
are limited to the corresponding contributions, Eq.~(\ref{source}),
of a maximally strong external fluid flow during the simulations.

At a physical length scale $r$ at
cosmic time $t$ the accuracy requirement on the step-size
is~\cite{recipes}
$\Delta\ln T\lesssim r^2/\left[t\,{\rm Max}(\kappa,\alpha)\right]$.
For about $10^4$ time steps per decade in $T$ and for the
coefficients given by Eq.~(\ref{coeff}) this is typically
fulfilled for temperatures up to close to the GUT scale
and co-moving lengths down to the parsec scale which are
mostly of interest here. Although
this accuracy requirement is not fulfilled at the smallest
length scales close to the inverse temperature used in
the numerical integration, the implicit method assures
at least convergence toward the equilibrium solution
at such scales.

The power spectra $b^2(k)$ and $h(k)$ can be obtained
by inverting the transformations of Eq.~(\ref{kspace}),
but a rough estimate is given by $b^2(k)\sim M_L(1/k)$
and $h(k)\sim H(1/k)/k^2$.

In the following we parametrize the magnetic seed field
by
\begin{equation}
  M_L(r)=N\frac{8\pi^3}{90}\frac{1}{(1+r/r_B)^n}\,,\label{iniM_L}
\end{equation}
where $N$ characterizes the strength relative to thermal
density, $r_B$ is the scale on which it is concentrated, and
$n$ is the power law index at much larger scales (causally
produced fields correspond to $n\geq5$).

\begin{figure}[ht]
\includegraphics[width=0.48\textwidth,clip=true]{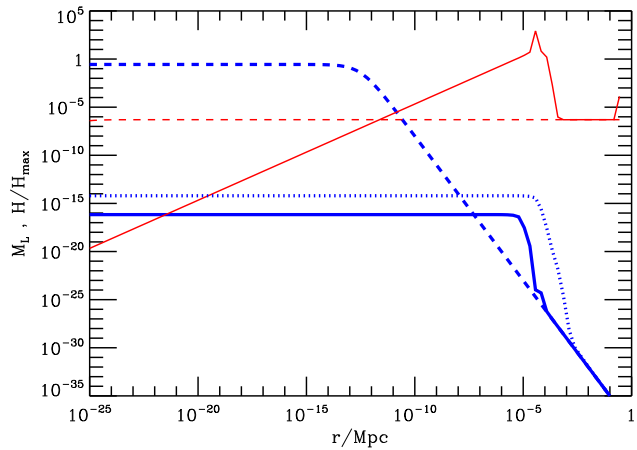}
\caption[...]{\label{fig1} Results in co-moving length scale $r$
for the case without external
fluid flow. Thick lines show $M_L(r)$ in units of $T^4$, for
initial condition at $T=100\,$GeV (thick dashed), and at zero
redshift (thick solid, i.e. $M_L=1$ corresponds to a
field strength of $\simeq1.4\times10^{-6}$ Gauss today,
note that $M_L$ is quadratic in ${\bf B}$). Thin lines show
helicity relative to maximal, $h(r)\equiv H(r)/H_{\rm max}(r)$, for
initial condition (thin dashed), and at zero redshift (thin solid).
For comparison, the thin dotted line shows the final $M_L(r)$
for maximal initial helicity (not shown).}
\end{figure}

To demonstrate the general effect of helicity we start
with magnetic fields of non-vanishing helicity in the
absence of source terms. We start at the electroweak scale,
$T=100\,$GeV, with a seed field
Eq.~(\ref{iniM_L}) with $N=0.1$, concentrated
at a scale $r_B=10^{-4}t_u(100\,{\rm GeV})$,
and a power law index $n=3$, motivated by a superposition of
magnetic dipoles, as may be expected for the electroweak
transition~\cite{soj}. We assume an initial helicity
$H_M\sim100n_b$, somewhat larger than the baryon number $n_b$,
as suggested by models~\cite{cornwall,vachaspati}. Assuming
the relative helicity, $h(r)\equiv H(r)/H_{\rm max}(r)$,
to be roughly independent of $r$, this corresponds to
$H(r)\sim100/N(n_b/n_\gamma)\,M_L(r)/r
\simeq5\times10^{-7}H_{\rm max}(r)$,
where the baryon to photon ratio $n_b/n_\gamma\sim5\times10^{-10}$.
Fig.~\ref{fig1} shows results for $M_L$ and the relative
helicity. The field at zero redshift is decreased
by dissipation up to the $\simeq0.1\,$parsec scale, whereas inverse
cascades enhance the field on scales of a few parsecs, reaching
$\sim10^{-14}\,$G. For comparison Fig.~\ref{fig1}
also shows the larger enhancement of $M_L$ for maximal initial
helicity (the case discussed in Ref.~\cite{fc}) which, however,
we consider speculative in the
absence of a specific model predicting such large helicities.

It is easy to show that the total helicity $H_M$ which
is dominated by the peak of $h(r)$ in Fig.~\ref{fig1} is
roughly conserved, and thus evolution is dominated by magnetic back-reaction
onto the fluid. Indeed, conservation of $H_M$ is usually employed
to estimate the field strength via $B^2\sim H_M/l_c$ which requires
an analytical estimate of the coherence scale $l_c$~\cite{vachaspati}.
In our numerical approach $l_c$ comes out without further assumptions
as the scale where the correlation function cuts off.

\begin{figure}[ht]
\includegraphics[width=0.48\textwidth,clip=true]{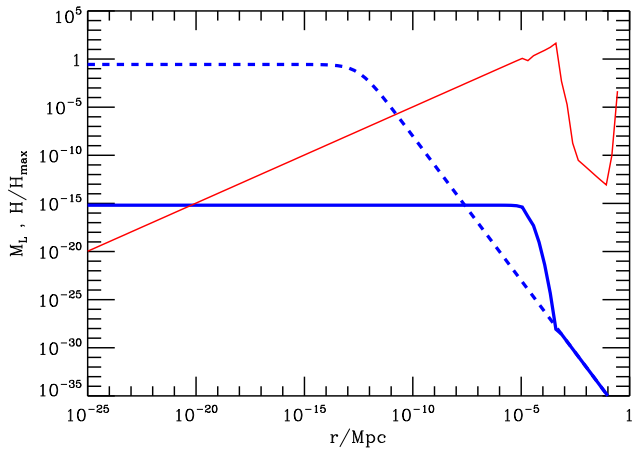}
\caption[...]{\label{fig2} Results in the presence of a fully
helical external fluid flow of small amplitude,
$\left\langle v_e^2\right\rangle=10^{-12}$, at the electroweak transition.
See Fig.~\ref{fig1} for line key and text for details.}
\end{figure}

Another interesting situation could be the production
of baryon and lepton number comparable to unity, $n_b/n_\gamma\sim1$
at $T=T_n\gg100\,$GeV, for instance during a phase transition related
to new physics, which could give rise to maximally helical flows as well.
These flows would consist of the tightly coupled electroweak plasma
and could survive as a small perturbation at least down to the neutrino
decoupling temperature, i.e. $t_{\rm corr}\simeq t_u(T)$ for $T\gtrsim\,$MeV.
Their amplitude can be estimated by the dilution factor
$\left\langle v_e^2\right\rangle\sim(n_b/n_\gamma)_{\rm today}^{4/3}
\sim10^{-12}$ due to the necessary entropy production above the
electroweak scale. Assuming a causal flow with power on scales not
too far below the horizon scale, we use $f(r)=[1+rT/t_u(T_n)/T_n]^{-5}$
with $T_n=10^{13}\,$GeV for the other parameters in Eq.~(\ref{vparam}).
We start with the same magnetic field produced at the electroweak
transition as above, but this time with vanishing initial helicity,
$H(r)\equiv0$. Fig.~\ref{fig2} shows that in this case the magnetic
field develops helicity and reaches values close to
$10^{-13}\,$G up to about 100 parsecs. The coherence scales are
also consistent with analytical estimates~\cite{son,fc}, but
are considerably larger than in Ref.~\cite{vachaspati}.

Our results also demonstrate that the presence of helicity
prevents complete dissipation of the fields at small scales,
resulting in a flat correlation function up to the coherence scale.
Furthermore, the relative magnetic helicity rises linearly with
$r$ and is close to maximal at the coherence scale.
This could have ramifications for the actual detection of helicity, for
example, via its effects on the CMB~\cite{pvw} and could be
an important signature of physics at or above the electroweak
scale.

\section{Conclusions}
We used the evolution equations for the two-point correlation
function of helical magnetic fields in MHD approximation
including magnetic diffusion, fluid viscosity, and back-reaction
onto the external fluid to evolve weakly helical fields produced
in the early Universe up to today. We find that
magnetic fields and/or fluid flows with a helicity relative
to maximal not much larger than the baryon to
photon number $\sim10^{-9}$, as expected during the
electroweak period, can lead to significant inverse
cascades. This results in magnetic fields that can be enhanced
by several orders of magnitude compared to the merely red-shifted
and frozen-in initial
fields at scales in the parsec and kilo-parsec range today. If
the seeds are roughly thermal in strength and if their power
is concentrated on scales not much smaller than the horizon
scale around the electroweak transition, the coherence length
is close to a kilo-parsec with field strengths up
to $\,10^{-13}\,$G. While this is smaller than the analytical
estimates in Refs.~\cite{fc}, it is based on the more realistic assumptions
of small helicities of order the baryon to photon number
where fluid viscosity can not be neglected. The fields we obtain are
certainly larger than from ``astrophysical'' seed field
mechanisms such as the Biermann battery, but are also
well within the limits from big bang nucleosynthesis and
the CMB (the best of which are $\sim10^{-9}\,$Gauss
on kpc--Mpc scales, see, e.g.,~\cite{lk,sb,jko}), and from
gravity wave production ($B(r)\lesssim10^{-11}(r/100\,{\rm pc})^{-3}$
for $n=3$~\cite{cd}). Such fields may also be testable by
ultra-high energy cosmic ray deflection~\cite{bs}.
The approach presented here can also be applied to other
magneto-genesis scenarios with pseudo-scalar seeds such as in
string cosmology~\cite{ss} where coupling to axions may lead
to larger helicities.

{\it Acknowledgments} I would like to thank A.~Buonanno, K.~Jedamzik, and
M.~Sakellariadou for helpful discussions.

\end{document}